\documentclass[twocolumn,showpacs,preprintnumbers,aps,nofootinbib]{revtex4-1}
\usepackage{graphicx}
\usepackage{dcolumn}
\usepackage{amsmath,amssymb,amsfonts}
\usepackage{latexsym}
\usepackage{bbm}
\usepackage{color}
\begin{document}

\preprint{CTPU-PTC-18-33}

\title{Radiative Corrections to Triple Higgs Coupling and Electroweak Phase Transition: Beyond One-loop Analysis}

\author{Eibun Senaha}
\email{senaha@ibs.re.kr}
\affiliation{Center for Theoretical Physics of the Universe, Institute for Basic Science (IBS), Daejeon 34126, Korea}
\bigskip

\date{\today}

\begin{abstract}
We evaluate dominant two-loop corrections to the triple Higgs coupling and strength of a first-order electroweak phase transition in the inert Higgs doublet model.
It is found that sunset diagrams can predominantly enhance the former and reduce the latter. 
As a result, the triple Higgs coupling normalized by the standard model value at two-loop level is more
enhanced than the corresponding one-loop value.
\end{abstract}



\maketitle

\section{Introduction}
Higgs mechanism is one of the fundamental footings of the standard model (SM).
Though the new scalar particle with a mass of 125 GeV was discovered at the Large Hadron Collider (LHC)~\cite{125h}, 
its roles as the mass giver for all the SM particles and the symmetry breaker for 
$\text{SU}(2)_L\times \text{U}(1)_Y$ have not been fully established yet.
For the latter, measurement of a triple Higgs boson coupling ($\lambda_{hhh}$) is particularly of importance 
since it can exist only after the electroweak symmetry is spontaneously broken.

Information of $\lambda_{hhh}$ can be extracted from double Higgs production processes at colliders.
Current LHC data put the upper limit on
$\sigma(pp\to hh)\times\text{Br}(hh\to b\bar{b}\gamma\gamma)$, which constrains the $\lambda_{hhh}$
normalized by the SM value ($\kappa_\lambda$) as
$-8.2<\kappa_\lambda<13.2$ at ATLAS~\cite{Aaboud:2018ftw} 
and $-11<\kappa_\lambda <17$ at CMS~\cite{Sirunyan:2018iwt} at 95\% confidence level, respectively.

In Ref.~\cite{KOSY}, 
$\lambda_{hhh}$ is calculated at one-loop level 
in a softly $Z_2$-broken two Higgs doublet model (2HDM). 
It is found that the extra heavy Higgs boson loop corrections can significantly enhance $\lambda_{hhh}$,
As a result, $\kappa_\lambda$ can be $\mathcal{O}(2\mbox{-}4)$.
This can occur if the one-loop corrections of the heavy Higgs bosons have the nondecoupling properties,
{\it i.e.}, the power corrections such that $(\text{heavy Higgs mass})^4$.

As found in Refs.~\cite{Grojean:2004xa,Kanemura:2004ch}, 
the sizable $\lambda_{hhh}$ would imply that the electroweak phase transition (EWPT) could be strongly first order as needed for successful electroweak baryogenesis~\cite{ewbg}. 
As is the enhancement of $\lambda_{hhh}$, the nondecoupling heavy Higgs bosons play a pivotal role,
and $\mathcal{O}(1)$ quartic couplings in the Higgs potential are the common drivers.
While the calculations are still within the weakly coupled regime satisfying a perturbative unitarity, 
quantification of the higher-order corrections are vitally important
for $\lambda_{hhh}$ measurements at future colliders. 
For instance, it is expected that $0.2<\kappa_\lambda<6.9$ at the high luminosity LHC with an integrated luminosity of $3~\text{ab}^{-1}$~\cite{D.Delgove}.
At the International Linear Collider, on the other hand,
the $27\%$ accuracy would be reachable with the full data set of 250+500 GeV~\cite{Fujii:2015jha}. 
Moreover, 100 TeV hadron colliders with an integrated luminosity of 30 $\text{ab}^{-1}$ can attain $(5\text{-}7)\%$ accuracy~\cite{3h_100TeV}. 

In this paper, we quantity the dominant two-loop contributions both $\lambda_{hhh}$ and 
strength of the first-order EWPT and clarify to what extent the correlation between them could be modified
compared to the one-loop result. 
As an illustration, we consider the inert two Higgs doublet model (IDM), which is also motivated by
dark matter (DM) physics. 
We point out that the magnitude of the nondecoupling effect is rather restricted 
by a vacuum condition associated with the DM phenomenology. 
However, the modification by the two-loop contributions is still relevant to the correlation between 
$\kappa_\lambda$ and strength of the first-order EWPT.

\section{Calculation scheme}
We expand the effective $hhh$ vertex defined in the on-shell (OS) scheme in powers of momenta as
\begin{align}
\hat{\Gamma}_{hhh}(p_1^2,p_2^2,p_3^2) \simeq \hat{\Gamma}_{hhh}(0,0,0)+\cdots.
\label{Ghhh_ex}
\end{align}
Apart from a threshold enhancement that occurs when incoming momentum 
is twice as large as the masses of particles running in loops,
the dominant quantum contributions come from the momentum-independent part~\cite{KOSY}.
Moreover, since we are interested in a deviation of the effective $hhh$ vertex from the SM value 
in new physics models such as the IDM, the ratio of $\hat{\Gamma}^{\text{NP}}_{hhh}(p_1^2,p_2^2,p_3^2)/\hat{\Gamma}_{hhh}^{\text{SM}}(p_1^2,p_2^2,p_3^2)$ is well approximated by $\hat{\Gamma}_{hhh}^{\text{NP}}(0,0,0)/\hat{\Gamma}_{hhh}^{\text{SM}}(0,0,0)\equiv \kappa_\lambda$~\cite{KOSY}.
Therefore, we will exclusively focus on the momentum-independent term in Eq.~(\ref{Ghhh_ex}) 
in this paper. 
Calculation of $\hat{\Gamma}_{hhh}(0,0,0)$ is greatly simplified if an effective potential is used. 
Let us define $\hat{\lambda}_{hhh}$ as
\begin{align}
-\hat{\Gamma}_{hhh}(0,0,0) \equiv \hat{\lambda}_{hhh} = \hat{Z}_h^{3/2}\lambda_{hhh},
\end{align}
where $\lambda_{hhh}$ is the third derivative of the effective potential ($V_{\text{eff}}$) 
defined in the $\overline{\text{MS}}$ scheme
and $\hat{Z}_h=Z_h^{\text{OS}}/Z_h^{\overline{\text{MS}}}$ with $Z_h^{\text{OS}}$ 
being the Higgs wavefunction renormalization constant 
in the OS scheme and $Z_h^{\overline{\text{MS}}}$ in the $\overline{\text{MS}}$ scheme. 

Though $\lambda_{hhh}$ is calculated up to two-loop level in supersymmetric SMs~\cite{lam3h_2L}, 
the analytic expression of $\lambda_{hhh}$ in the SM seems absent in the literature. 
We thus start with the SM case using our calculation scheme.

The tree-level Higgs potential is given by
\begin{align}
V_0(\Phi) = -\mu_\Phi^2\Phi^\dagger \Phi+\lambda(\Phi^\dagger\Phi)^2,\quad
\Phi=
\begin{pmatrix}
G^+ \\
\frac{1}{\sqrt{2}}(v+h+iG^0)
\end{pmatrix},\label{SMHiggs}
\end{align}
where $v$ denotes the vacuum expectation value (VEV) of the Higgs boson ($h$) and $G^{0,\pm}$ are the Nambu-Goldstone bosons.
We calculate $\lambda_{hhh}$ using the $\overline{\text{MS}}$-regularized effective potential 
at two-loop level~\cite{Coleman:1973jx,Ford:1992pn}, which is expanded as
\begin{align}
V_{\text{eff}}(\varphi) = V_0(\varphi)+V_1(\varphi)+V_2(\varphi),
\end{align}
where $\varphi$ denotes the background classical field.
With this, the Higgs mass and the triple Higgs coupling are, respectively, defined as
\begin{align}
m_h^2 &= \frac{\partial^2 V_{\rm eff}}{\partial \varphi^2}\bigg|_{\varphi=v}
= 2\lambda v^2 + \mathcal{D}_m \Delta V_{\text{eff}}(\varphi),\label{mhSM} \\ 
\lambda_{hhh}^{\text{SM}}&=\frac{\partial^3 V_{\rm eff}}{\partial \varphi^3}\bigg|_{\varphi=v}
=\frac{3m_h^2}{v}+\mathcal{D}_\lambda \Delta V_{\text{eff}}(\varphi),\label{lam3hSM}
\end{align}
where 
\begin{align}
\mathcal{D}_m &= 
\left[
	\frac{\partial^2}{\partial\varphi^2}-\frac{1}{v}\frac{\partial }{\partial \varphi}
\right]_{\varphi=v}, \\
\mathcal{D}_\lambda &= 
\left[\frac{\partial^3}{\partial\varphi^3}-\frac{3}{v}
\left(
	\frac{\partial^2}{\partial\varphi^2}-\frac{1}{v}\frac{\partial }{\partial \varphi}
\right)\right]_{\varphi=v},
\end{align}
with $\Delta V_{\text{eff}}(\varphi)=V_1(\varphi)+V_2(\varphi)$.
Note that $\mu_\Phi$ is eliminated by a minimization condition so that $m_h^2$ is defined in the minimum of $V_{\text{eff}}$. Furthermore, since $\lambda$ is replaced by $m_h^2$ using Eq.~(\ref{mhSM}),
the leading log corrections at each loop in Eq.~(\ref{lam3hSM}) are absorbed by the Higgs mass renormalization as explicitly demonstrated below.

\begin{figure}[t]
\center
\includegraphics[width=2.0cm]{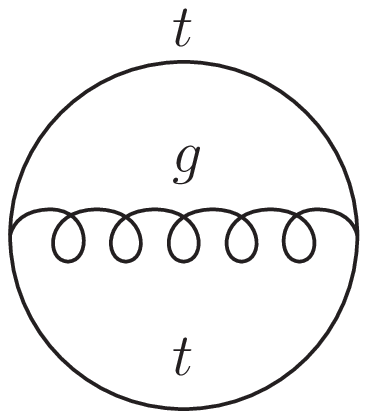}
\hspace{0.5cm}
\includegraphics[width=2.0cm]{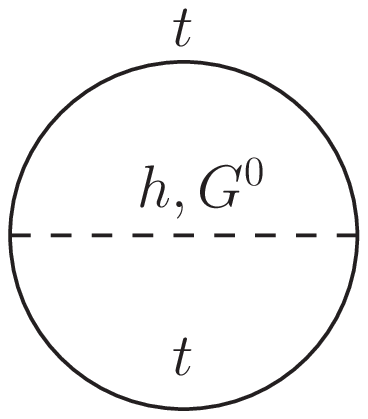}
\hspace{0.5cm}
\includegraphics[width=2.0cm]{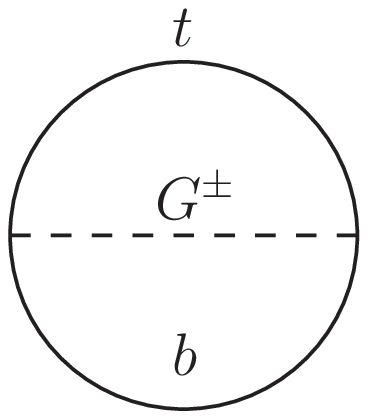}
\caption{Two-loop diagrams contributing to $\lambda_{hhh}^{\text{SM}}$ to $\mathcal{O}(g_3^2y_t^4)$ and $\mathcal{O}(y_t^6)$.}
\label{fig:Veff_2L_SM}
\end{figure}

The dominant two-loop contributions arise from the sunset diagrams as depicted
in Fig.~\ref{fig:Veff_2L_SM}.
From them, one can obtain the $\mathcal{O}(g_3^2y_t^4)$ 
and $\mathcal{O}(y_t^6)$ corrections to $\lambda_{hhh}^{\text{SM}}$,
where $g_3$ and $y_t$ are the SU(3$)_C$ and top Yukawa couplings, respectively.
Combining the dominant one-loop contribution coming from the top quark, one finds
\begin{align}
\lambda_{hhh}^{\text{SM}} &= \frac{3m_h^2}{v}+\Delta^{(1)}\lambda_{hhh}^{\text{SM}}
+\Delta^{(2)}\lambda_{hhh}^{\text{SM}},
\end{align}
where
\begin{align}
\Delta^{(1)}\lambda_{hhh}^{\text{SM}} &=
\frac{1}{16\pi^2}\left(-\frac{48m_t^4}{v^3}\right), \\
\Delta^{(2)}\lambda_{hhh}^{\text{SM}} &=
\frac{1}{(16\pi^2)^2}\frac{m_t^4}{v^3}
\bigg[
768g_3^2\left(\ell_t+\frac{1}{6}\right) \nonumber\\
&\hspace{3cm}	
	-144y_t^2\left(\ell_t-\frac{7}{6}\right)
\bigg],
\end{align}
with $m_t=y_t v/\sqrt{2}$ and $\ell_t=\ln(m_t^2/\bar{\mu}^2)$ with $\bar{\mu}$ being the renormalization scale. 
As mentioned above, the leading-log terms at each loop level
are absorbed by the $m_h$ renormalization.
As a result, the one-loop leading contribution becomes $\mathcal{O}(m_t^4)$~\cite{Hollik:2001px}.
Likewise, after absorbing the double-log terms at two-loop level, 
one has $\mathcal{O}(m_t^4)$ with extra coefficients including the single log terms.
Noting that all the parameters appearing in $\lambda_{hhh}^{\text{SM}}$ are the $\overline{\text{MS}}$ running parameters, the $\ell_t$ terms at two-loop level
can be absorbed into $m_t$ at one-loop order
using the renormalization group (RG) equations.
After expressing all the $\overline{\text{MS}}$ variables evaluated at the pole top quark mass ($M_t=173.1$ GeV~\cite{Tanabashi:2018oca})
with the physical ones retaining the $\mathcal{O}(g_3^2)$ and $\mathcal{O}(y_t^2)$ terms
(for explicit expressions, see, {\it e.g.}, Eqs.~(2.18) and (2.19) in Ref.~\cite{Degrassi:2012ry}),
one arrives at
\begin{align}
\hat{\lambda}_{hhh}^{\text{SM}}
&\simeq  \frac{3M_h^2}{v_{\text{phys}}}
\Bigg[
	1+\frac{1}{16\pi^2}\left(-\frac{16M_t^4}{M_h^2v_{\text{phys}}^2} + \frac{7}{2}\frac{M_t^2}{v_{\text{phys}}^2}\right) \nonumber\\
&\hspace{1.5cm}	
	+\frac{1}{(16\pi^2)^2}\frac{M_t^4}{M_h^2v_{\text{phys}}^2}
	\left(384g_3^2-\frac{312M_t^2}{v_{\text{phys}}^2}\right)
\Bigg] \nonumber\\
&=(190.38~\text{GeV})\times \Big[1-8.5\%+1.4\%\Big]=176.85~\text{GeV},
\label{lam3h_SM}
\end{align}
where $v_{\text{phys}}^2=1/(\sqrt{2}G_F)$
with $G_F(=1.166\times 10^{-5}~\text{GeV}^{-2})$ being the Fermi coupling constant, $M_h=125.0$ GeV is the pole mass of the Higgs boson and $g_3(M_t)=1.167$. 
One can see that $\lambda_{hhh}$ gets enhanced compared to the one-loop result.
Note that the additional one-loop correction arises when converting the $\overline{\text{MS}}$ parameters into 
the OS ones, which has the $+1.1\%$ contribution. \footnote{The omitted terms such as the gauge boson contributions amount to about +0.6\% contribution.} Even though it is subleading at one loop order, it is comparable to the two-loop corrections so that it is not negligible. 
In our numerical study, we also take the leading one-loop corrections of the gauge and Higgs bosons into account using the effective potential method.
Our numerical calculations show that $\hat{\lambda}_{hhh}^{\text{SM}}=176.23$ GeV and $180.24$ GeV at one and two-loop levels, respectively~\cite{ES}.
We have checked that the corresponding one-loop value of H-COUP~\cite{Kanemura:2017gbi} is $\hat{\lambda}_{hhh}^{\text{SM}}=178.01$ GeV, so the relative error is 0.9\%. The difference may come from subleading gauge bosons contributions that are missing in our calculation. 
Such contributions are not sufficiently small compared to the dominant two-loop contributions evaluated above
so that they have to be included in the full calculation of $\hat{\lambda}_{hhh}^{\text{SM}}$. 
However, this omission has little effect on $\kappa_\lambda$ in the IDM due to sizable new physics effects as discussed below.

\section{Model}
As a benchmark model, we consider the IDM
in which a $Z_2$-odd Higgs doublet ($\eta$) is added to the SM~\cite{Barbieri:2006dq,IDM,Kanemura:2016sos,Belyaev:2016lok}.
It is known that the model can accommodate both the strong first-order EWPT and the dark matter (DM) candidate simultaneously~\cite{IDM_EWPT,Blinov:2015vma,Laine:2017hdk}.
We quantify the leading two-loop corrections of the extra Higgs bosons to $\lambda_{hhh}$
in a cosmologically interesting region.

As a result of the $Z_2$ parity ($\Phi\to \Phi$ and $\eta\to -\eta$),
the Higgs potential is cast into the form
\begin{align}
\lefteqn{V_0^{\text{IDM}}(\Phi, \eta)}\nonumber\\
& = \mu_1^2\Phi^\dagger \Phi+\mu_2^2\eta^\dagger \eta
+\frac{\lambda_1}{2}(\Phi^\dagger \Phi)^2
+\frac{\lambda_2}{2}(\eta^\dagger \eta)^2 \nonumber\\
&\quad 
+\lambda_3(\Phi^\dagger \Phi)(\eta^\dagger \eta)
+\lambda_4 (\Phi^\dagger \eta)(\eta^\dagger \Phi)
+\frac{\lambda_5}{2}\Big[(\Phi^\dagger \eta)^2+\text{H.c}\Big],
\end{align}
where $\Phi$ is the same as in the SM given in Eq.~(\ref{SMHiggs}) and $\eta$ is parametrized as
\begin{align}
\eta = 
\begin{pmatrix}
H^+ \\
\frac{1}{\sqrt{2}}(H+iA)
\end{pmatrix}.
\end{align}
The Higgs boson masses are expressed at tree level as
$m_h^2  =\lambda_1v^2$ and $m_\phi^2 = \mu_2^2+\bar{\lambda}_{h\phi\phi}v^2/2$ for $\phi=H,A,H^\pm$, 
where $\bar{\lambda}_{hHH}=\lambda_3+\lambda_4+\lambda_5$, 
$\bar{\lambda}_{hAA}=\lambda_3+\lambda_4-\lambda_5$, and 
$\bar{\lambda}_{hH^+H^-}=\lambda_3$. 
In our analysis, $H$ is assumed to be the DM.
The pole Higgs masses are denoted as $M_h$, $M_H$, $M_A$ and $M_{H^\pm}$,
respectively, and we trade $\{\mu_1^2, \mu_2^2, \lambda_1, \lambda_2, \lambda_3, \lambda_4, \lambda_5\}$ with $\{v_{\text{phys}}, M_h, M_H, M_A, M_{H^\pm},\bar{\lambda}_{hHH},\lambda_2\}$
as the input parameter set. 

It is easy to obtain the one-loop contributions of the extra particles
to $\lambda_{hhh}$, which takes the form~\cite{Kanemura:2016sos}
\begin{align}
\Delta^{(1)}\lambda_{hhh}^{\text{IDM}} 
& = \sum_{\phi=H,A,H^\pm}
	n_\phi\frac{ 4m_\phi^4}{16\pi^2v^3}\left(1-\frac{\mu_2^2}{m_\phi^2}\right)^3,
\label{lam3h_IDM1L}
\end{align}
where $n_H=n_A=1$ and $n_{H^\pm}=2$.
As found in Refs.~\cite{KOSY,Kanemura:2016sos}, the one-loop correction can grow with $m_\phi^4$
if $\mu_2^2\ll m_\phi^2$, {\it i.e.}, $m_\phi^2\simeq \bar{\lambda}_{h\phi\phi} v^2/2$ ({\it nondecoupling regime}).
In the opposite limit of $m_\phi^2\simeq \mu_2^2$, on the other hand, $\Delta^{(1)}\lambda_{hhh}^{\text{IDM}}$ would be suppressed in the large $m_\phi$ limit ({\it decoupling regime}). 
As discussed below, the nondecoupling regime is exactly the condition that EWPT is strongly first order. 
In our work, we consider $M_H\simeq M_h/2$
so that the former limit applies only for $A$ and $H^\pm$. 
Furthermore, $M_A=M_{H^\pm}$ is taken to avoid the $\rho$ parameter constraint~\cite{Barbieri:2006dq}. 
In this case, one has $\bar{\lambda}_{hAA}=\bar{\lambda}_{hH^+H^-}=\lambda_3$. 

At two-loop order, the dominant corrections to $\lambda_{hhh}$ come from the sunset diagrams with
$\mathcal{O}(\lambda_3^3)$ in magnitude.
For illustrative purposes, we consider the case of $m_h\ll m_A, m_{H^\pm}$ in the nondecoupling regime.
In this limit, $m_h$ can be considered as a small perturbation 
and hence the $\mathcal{O}(\lambda_3^3)$ contributions have the simple form
\begin{align}
&\Delta^{(2)} \lambda_{hhh}^{\text{IDM}}
\simeq \sum_{\phi=A,H^\pm}\frac{8n_\phi\bar{\lambda}_{h\phi\phi}^2}{(16\pi^2)^2}\frac{m_\phi^2}{v}
\left(\ell_\phi-\frac{1}{2} \right),
\end{align}
where $\ell_\phi=\ln(m_\phi^2/\bar{\mu}^2)$.
As is the top quark contribution, the log terms are absorbed into $m_\phi^2$ in Eq.~(\ref{lam3h_IDM1L})
by use of the RG equations. 
Putting all together, one arrives at
\begin{align}
\lefteqn{\Delta^{(1)}\lambda_{hhh}^{\text{IDM}} +\Delta^{(2)}\lambda_{hhh}^{\text{IDM}}}\nonumber\\
&\simeq \sum_{\phi=A,H^\pm}\frac{4n_\phi}{16\pi^2v^3}
\bigg[
m_\phi^4(m_\phi)-\frac{4m_\phi^6}{16\pi^2v^2}
\bigg]+\cdots,\label{lam3h_2L_analy}
\end{align}
where $m_\phi(m_\phi)$ are the $\overline{\text{MS}}$-running masses of $\phi$ evaluated at $m_\phi$.
We choose $M_Z=91.1876$ GeV as the input scale for all the running parameters
as in Refs.~\cite{Blinov:2015vma,Laine:2017hdk}.  
Modification due to the two-loop contributions mainly comes through the RG running effects 
in the first term, which enhances $\lambda_{hhh}$.

It is found that the $\mathcal{O}\big(M_h^2M_\phi^2/(16\pi^2v_{\text{phys}}^3)\big)$ terms appear
when expressing $\hat{\lambda}_{hhh}^{\text{IDM}}$ with the physical parameters. 
However, they are accidentally cancelled for $M_A=M_{H^\pm}$.
Use of the analytic expression (\ref{lam3h_2L_analy}) yields overestimated results for nonzero $\mu_2^2$
so that we evaluate $\Delta^{(2)}\lambda_{hhh}^{\text{IDM}}$ numerically for our quantitative studies~\cite{ES}.

\begin{figure}[t]
\center
\includegraphics[width=0.8\linewidth]{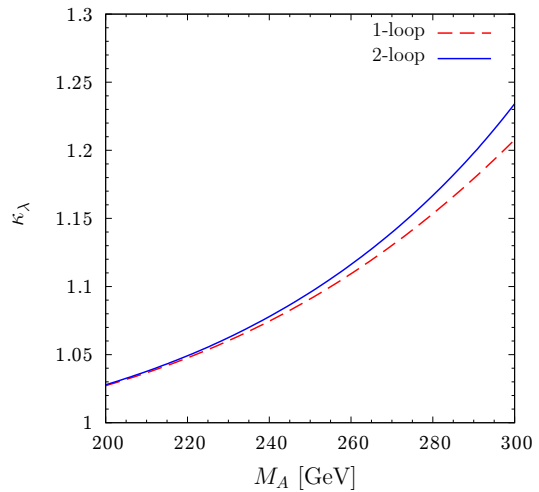}
\caption{$\kappa_\lambda$ as a function of $M_A$ at one and two loop levels denoted as the red-dashed and blue-solid curves, respectively. We take $M_H=62.7$ GeV, $\lambda_2=0.02$ and $\bar{\lambda}_{hHH}=4.6\times 10^{-3}$ at the $M_Z$ scale.}
\label{fig:kaph}
\end{figure}

In Fig.~\ref{fig:kaph}, $\kappa_\lambda$ at one and two-loop levels are plotted as functions of $M_A$
with the red-dashed and solid-blue curves, respectively. 
We take the two-loop value of $\hat{\lambda}_{hhh}^{\text{SM}}$ as the SM normalization in the both cases
and $M_H=62.7$ GeV, $\lambda_2(M_Z)=0.02$ and 
$\bar{\lambda}_{hHH}(M_Z)=4.6\times 10^{-3}$ as a benchmark. 
We find that using MicrOMEGAs~\cite{micromegas} the above parameter set
gives the DM relic density and the spin-independent
cross section of the DM with a proton as $\Omega_{\text{DM}}h^2=0.113$ and $\sigma_{\text{SI}}^p=4.6\times 10^{-47}~\text{cm}^2$, respectively, which are consistent with the current DM data~\cite{Tanabashi:2018oca,Aprile:2018dbl}.
For the LHC constraints,  comprehensive studies can be found in Ref.~\cite{Belyaev:2016lok}.
In our chosen parameter space, the most stringent constraint comes from the signal strength of the Higgs boson decays to two photons ($\mu_{\gamma\gamma}$) that can be affected by the charged Higgs bosons.
In our case, it is found that $\mu_{\gamma\gamma}\simeq 0.9$ for $200~\text{GeV}\le M_{H^\pm}\le300~\text{GeV}$ which is consistent with the current data $\mu_{\gamma\gamma}^{\text{ATLAS}}=0.99_{-0.14}^{+0.15}$~\cite{Aaboud:2018xdt} and $\mu_{\gamma\gamma}^{\text{CMS}}=1.18_{-0.14}^{+0.17}$~\cite{Sirunyan:2018ouh} within 2$\sigma$.

Since the one-loop corrections to $m_H$ is positive, 
$\mu_2^2$ has to be made smaller as $M_A$ increases in order to have $M_H=62.7$ GeV, 
and thus the nondecoupling effects become more pronounced as discussed above. 
\footnote{For $M_A=300$ GeV, one obtains $\lambda_3(M_Z)=2.82$ and 
$\lambda_4(M_Z)=\lambda_5(M_Z)=-1.41$. 
If we run those couplings using one-loop RG equations~\cite{Branco:2011iw},
one finds $\lambda_3(\bar{\mu})>4\pi$ at $\bar{\mu}\simeq 53.9$ TeV.}
As pointed out in Ref.~\cite{Blinov:2015vma}, however, that $M_A$ cannot be greater than a certain value due to the occurrence of $\mu_2^2<0$ that generates a nontrivial minimum along the inert doublet field direction,
which could be deeper than the prescribed electroweak vacuum and thus excluded.
In the chosen parameter set, it is found that $\mu_2^2\lesssim0$ for $M_A\gtrsim 300$ GeV.
Within the allowed range, $\kappa_\lambda$ at two-loop level can be enhanced up to about $2\%$. 
It should be noted, however, that the further enhancement could be possible if the requirement of $\mu_2^2>0$ were absence.
In the ordinary 2HDM, for instance, $\lambda_{hhh}$ can receive $\mathcal{O}(100)\%$ corrections
at one-loop level with increasing $M_A$ as mentioned above.
In this case, one may ask whether the power correction of the $m_\phi^6$ term in Eq.~(\ref{lam3h_2L_analy}) can compete with the one-loop ones.
Here, we give a simple argument that it would not happen. 
On the grounds of the dimensional analysis, 
the dominant power corrections to $\lambda_{hhh}$ at $\ell$-loop order may be cast into the form
\begin{align}
\Delta^{(\ell)}\lambda_{hhh}\sim (-1)^{\ell+1}\frac{m_\phi^2}{v}\left(\frac{4m_\phi^2}{16\pi^2 v^2}\right)^\ell,
\end{align}
where $\mu_2^2=0$ and combinatorial factors are ignored. 
If the expansion parameter $m_\phi^2/(4\pi^2 v^2)$ is close to unity,
one obtains $m_\phi\simeq 2\pi v=1546$ GeV which corresponds to $\bar{\lambda}_{h\phi\phi}\simeq 8\pi^2$. This is clearly far beyond the perturbativity bound. Conversely, if we require $\bar{\lambda}_{h\phi\phi}=2m_\phi^2/v^2<4\pi$ as a crude perturbativity criterion, $m_\phi^2/(4\pi^2 v^2)<1/(2\pi)\simeq 0.16$. Thus, the maximal two-loop power corrections amount to about $-16\%$ of the one-loop ones. 

\begin{figure*}[t]
\center
\includegraphics[width=0.4\linewidth]{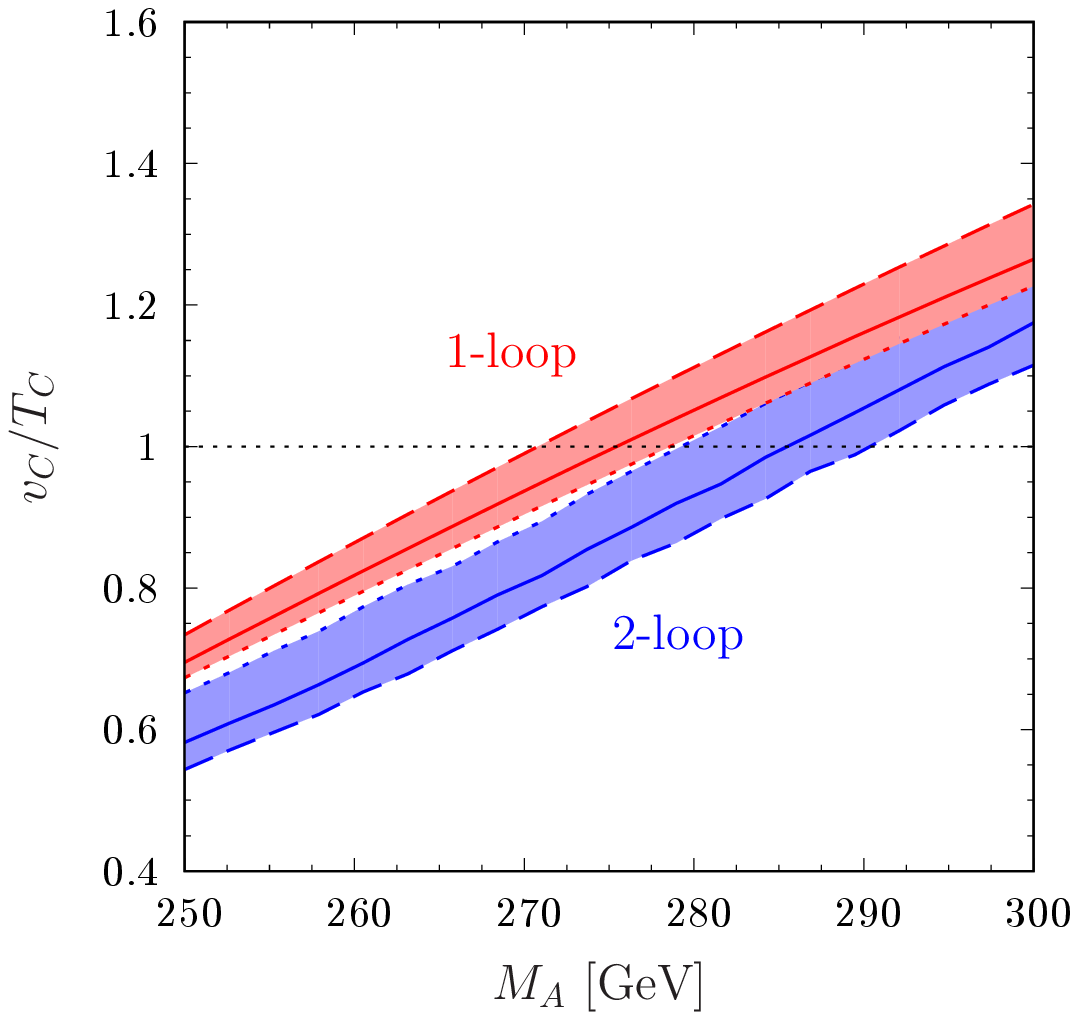}
\hspace{0.5cm}
\includegraphics[width=0.4\linewidth]{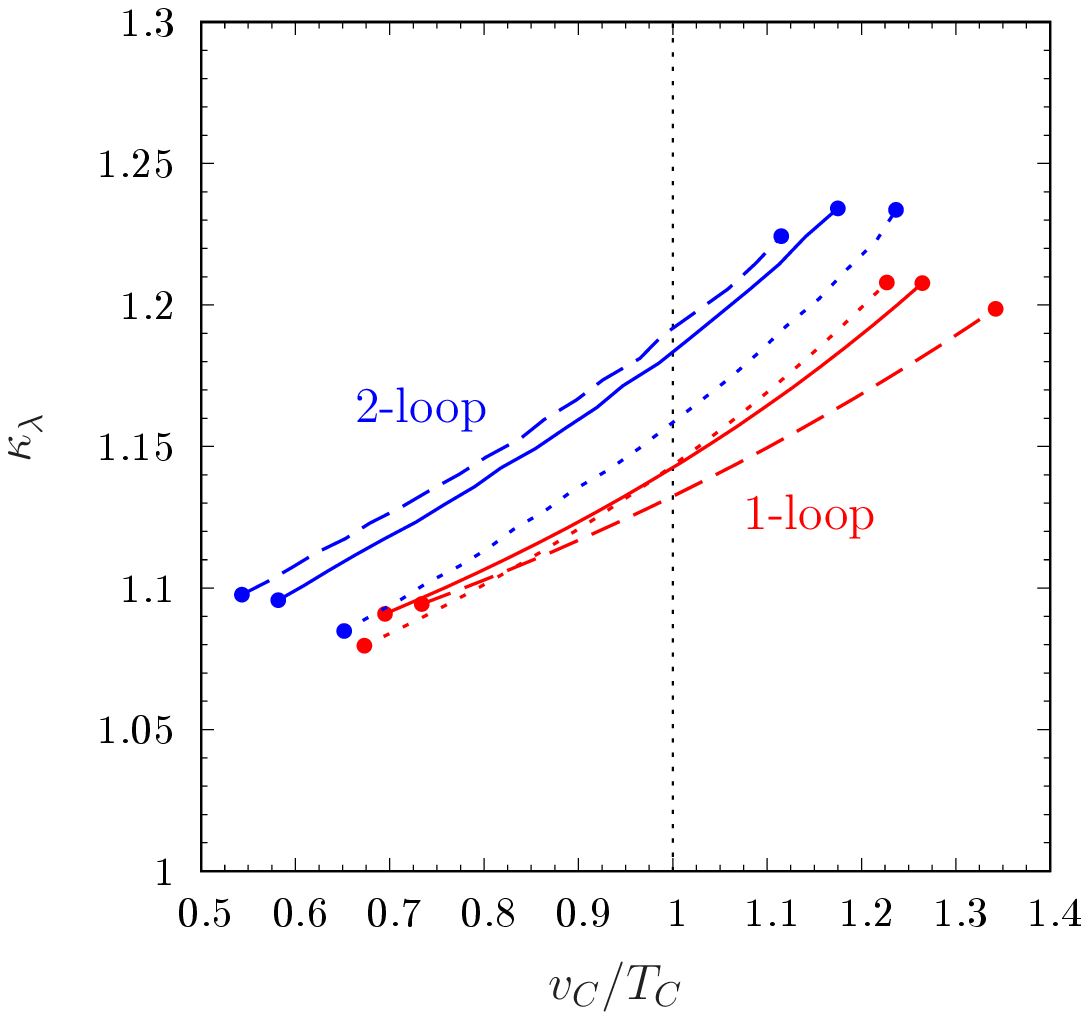}
\caption{(Left) Shown is the plot of $v_C/T_C$ against $M_A$ at one (red-band) and two-loop (blue-band) levels varying $\bar{\mu}$ from $0.5M_A$ (dotted corves) to $1.5M_A$ (dashed curves), and the solid curves correspond to $\bar{\mu}=M_A$.
(Right) $v_C/T_C$-$\kappa_\lambda$ correlations at one and two-loop levels varying $M_A$ from 250 GeV to 300 GeV, where the style and color schemes of the curves are the same as those in the left panel. The input parameters are the same as in Fig.~\ref{fig:kaph}.}
\label{fig:lam3hEWPT}
\end{figure*}

\section{$\lambda_{hhh}$-EWPT correlation}
It is known that the remnant of the strong first-order EWPT can appear  in $\lambda_{hhh}$.
Before conducting the two-loop analysis, we briefly outline the $\lambda_{hhh}$-EWPT correlation 
at one-loop.
The criterion for the strong first-order EWPT is given by~\cite{ewbg}
\begin{align}
\frac{v_C}{T_C}>\zeta_{\text{sph}}\label{BNPC},
\end{align}
where $T_C$ is a temperature at which there are two degenerate vacua in the effective potential,
$v_C$ is the Higgs VEV at $T_C$, and $\zeta_{\text{sph}}$ depends on the sphaleron profile etc,
and typically, $\zeta_{\text{sph}}\simeq 1$. 
Use of the high-$T$ expansion (HTE) of the one-loop thermal function~\cite{Dolan:1973qd}
makes it easy to see the $\lambda_{hhh}$-EWPT correlation. At $T_C$ the effective potential is cast into the form
\begin{align}
V_{\text{eff}}(\varphi; T_C) = \frac{\lambda_{T_C}}{4}\varphi^2(\varphi-v_C)^2,
\quad v_C=\frac{2ET_C}{\lambda_{T_C}},\label{1LHTE}
\end{align}
where $E$ denotes the coefficient of the $\varphi^3$ term. In the SM, $E_{\text{SM}}\simeq0.01$
coming from the gauge bosons.
In the IDM model, on the other hand, the extra Higgs bosons yield the contributions of $-T(\bar{m}_\phi^2)^{3/2}/(12\pi)$ in $V_{\text{eff}}(\varphi; T)$.
As is the gauge boson case, the $\varphi^3$ term can be generated if $\bar{m}_\phi^2\simeq \bar{\lambda}_{h\phi\phi}\varphi^2/2$, which contributes to $E$. Remarkably, this is exactly the case that $\Delta^{(1)}\lambda_{hhh}^{\text{IDM}}$ is enhanced, {\it i.e.}, nondecoupling regime. 
As mentioned above, only $A$ and $H^\pm$ can have such a limit and play an essential role in achieving the strong first-order EWPT. The additional contributions in $E$ are found to be
$\Delta E \simeq (m_A^3+2m_{H^\pm}^3)/(12\pi v^3)$.
One can find that the minimum values of $m_A$ and $m_{H^\pm}$ satisfying the criterion (\ref{BNPC}) 
sets the minimum deviation of $\lambda_{hhh}^{\text{IDM}}/\lambda_{hhh}^{\text{SM}}$.
In this way, the strong first-order EWPT inevitably leads to the significant deviation in $\lambda_{hhh}$.
Detailed knowledge of $\zeta_{\text{sph}}$ is of great importance in order
to quantify the amount of the deviation precisely (for an improvement of $\zeta_{\text{sph}}$ and its impact on $\lambda_{hhh}$ in the SM with a real singlet scalar, see Ref.~\cite{Fuyuto:2014yia}).

Now we extend the above discussion to two-loop level.
As far as the first-order EWPT is concerned, the sunset diagrams are more relevant than 
the figure-8 diagrams~\cite{VeffT2L}.  In the IDM, the relevant contributions are 
\begin{align}
V_2(\varphi; T)
&\ni -\frac{1}{4}\sum_{\phi=A,H^\pm} n_{\phi}
\Big[
	\lambda_{h\phi\phi}^2\bar{H}^{(T)}(\bar{m}_\phi^2,\bar{m}_\phi^2,\bar{m}_h^2)
\Big] \nonumber\\
&\simeq \sum_{\phi} n_\phi\frac{T^2\bar{\lambda}_{h\phi\phi}^2\varphi^2}{128\pi^2}
	 \ln\frac{\bar{m}_\phi^2}{T^2},\label{V2_HTE}
\end{align}
where $\bar{H}^{(T)}$ is the finite-temperature part of the sunset diagram~\cite{Arnold:1992rz}. 
In the second line, the HTE as well as $\bar{m}_h\simeq 0$ are assumed.
It is known that $\varphi^2\ln (\bar{m}^2/T^2)$ with positive (negative) coefficient would weaken (strengthen) the first-order EWPT~\cite{VeffT2L},
and the dominant scalar sunset diagrams in the IDM correspond to the former. 
From this simple argument,
one can infer that strength of the first-order EWPT would get smaller than those at one-loop level
while the other way around for $\lambda_{hhh}$.
In what follows, we evaluate the $\lambda_{hhh}$-EWPT correlation without using the HTE approximation
of Eq.(\ref{V2_HTE}) (details are given in a separate paper~\cite{ES}).

Following the thermal resummation and renormalization schemes adopted in Refs.~\cite{Tresum,Funakubo:2012qc}, 
we study $v_C/T_C$ numerically.
Previous two-loop analysis of EWPT in the IDM can be found in Ref.~\cite{Laine:2017hdk},
and our results are consistent with them within theoretical uncertainties
if we use their input parameters.

In the left panel of Fig.~\ref{fig:lam3hEWPT}, with the same input parameters as in Fig.~\ref{fig:kaph}, $v_C/T_C$ at one and two-loop levels against $M_A$ are shown as the red and blue bands respectively,
where we vary $\bar{\mu}$ from $0.5M_A$ to $1.5M_A$ with the solid, dashed and dotted curves being $\bar{\mu}=M_A, 1.5M_A$ and $0.5M_A$, respectively.

As expected from the qualitative discussion above, 
$v_C/T_C$ in both cases grow with increasing $M_A$ due to the nondecoupling effects
of $A$ and $H^\pm$. However, $v_C/T_C$ at two-loop level becomes smaller,
which is due mostly to the logarithmic terms in the sunset diagrams involving $A$ and $H^\pm$
as argued above.

Note that the $\bar{\mu}$ dependence shown here mostly arises from the $\ln(\bar{\mu}/T)$ terms 
in the thermal functions and it is not mitigated at two-loop level. 
To circumvent this issue, we may adopt nonperturbative approach or dimensionally-reduced effective field theory (for a recent study, see Ref.~\cite{Kainulainen:2019kyp}).
In spite of these uncertainties, it still holds that $v_C/T_C|_{\text{2-loop}}<v_C/T_C|_{\text{1-loop}}$.

In the right panel of Fig.~\ref{fig:lam3hEWPT}, the correlations between $v_C/T_C$ and $\kappa_\lambda$
are represented at one and two-loop levels. The style and color schemes of the curves are the same as those in the left panel. 
It is found that $\kappa_\lambda$ at two-loop level gets enhanced owing to the reduction of $v_C/T_C$.
Though the degree of modification varies with $\bar{\mu}$, 
the inequality of $\kappa_\lambda^{\text{1-loop}}<\kappa_\lambda^{\text{2-loop}}$ remains intact. 

\section{Conclusion and discussion}
We have quantified the two-loop effects on the triple Higgs coupling and strength
of the first-order EWPT in the IDM. We found that the sunset diagrams can alter both of the one-loop values.
As a result, $\kappa_\lambda$ is enhanced by about $+2\%$ while $v_C/T_C$ is reduced by some amount that varies with $\bar{\mu}$, for instance $(7\mbox{-}16)\%$ for $\bar{\mu}=M_A$.
The magnitudes of the corrections are restricted by the requirement of $\mu_2^2>0$.
Correspondingly, at two-loop level $M_A(=M_{H^\pm})$ is shifted upward by about 10 GeV and $\kappa_\lambda$  rises up to around 4\% in the region where EWPT is strongly first order.
We emphasize that even though the numerical values are fluctuated by the significant $\bar{\mu}$ dependence, it still holds that $\kappa_\lambda^{\text{1-loop}}<\kappa_\lambda^{\text{2-loop}}$.

We finally make a comment on a gauge dependence of $v_C/T_C$.
Our calculation method is not gauge invariant, 
and the Landau gauge is adopted~\cite{Arnold:1992rz}.
Since the dominant two-loop contributions are the scalar loops, their effects are not spoiled by the gauge artifact.
However, to solve this problem in addition to the aforementioned $\bar{\mu}$ dependence issue in a satisfactory manner, more refined calculation scheme is needed. We defer this to future work.


\begin{acknowledgments}
This work was supported by IBS under the project code, IBS-R018-D1.
\end{acknowledgments}

Note added.--- During the review process, we became aware of a similar work~\cite{Braathen:2019pxr}. We have confirmed that the coefficient of the $\mathcal{O}(M_t^6)$ term in Eq.~(\ref{lam3h_SM}) agrees with that of \cite{Braathen:2019pxr} after correcting a typo. Our findings of $\kappa_\lambda$ at two-loop level are consistent with theirs.



\begin{thebibliography}{99}

\bibitem{125h}  
  G.~Aad {\it et al.} [ATLAS Collaboration],
  Phys.\ Lett.\ B {\bf 716}, 1 (2012);~
%
  S.~Chatrchyan {\it et al.} [CMS Collaboration],
  Phys.\ Lett.\ B {\bf 716}, 30 (2012).

\bibitem{Aaboud:2018ftw} 
  M.~Aaboud {\it et al.} [ATLAS Collaboration],
  arXiv:1807.04873 [hep-ex].
%
\bibitem{Sirunyan:2018iwt} 
  A.~M.~Sirunyan {\it et al.} [CMS Collaboration],
  arXiv:1806.00408 [hep-ex].
  
\bibitem{KOSY}  
  S.~Kanemura, S.~Kiyoura, Y.~Okada, E.~Senaha and C.~P.~Yuan,
  Phys.\ Lett.\ B {\bf 558}, 157 (2003);~
%
  S.~Kanemura, Y.~Okada, E.~Senaha and C.-P.~Yuan,
  Phys.\ Rev.\ D {\bf 70}, 115002 (2004).
%

\bibitem{Grojean:2004xa} 
  C.~Grojean, G.~Servant and J.~D.~Wells,
  Phys.\ Rev.\ D {\bf 71}, 036001 (2005).

\bibitem{Kanemura:2004ch} 
  S.~Kanemura, Y.~Okada and E.~Senaha,
  Phys.\ Lett.\ B {\bf 606}, 361 (2005). 

\bibitem{ewbg}
  V.A.~Kuzmin, V.A.~Rubakov and M.E.~Shaposhnikov,
  Phys.\ Lett.\ B {\bf 155}, 36 (1985).
For some reviews, 
see e.g.
%
  M.~Quiros,
  Helv.\ Phys.\ Acta {\bf 67}, 451 (1994);
%
  V.A.~Rubakov and M.E.~Shaposhnikov,
  Usp.\ Fiz.\ Nauk {\bf 166}, 493 (1996)
  [Phys.\ Usp.\  {\bf 39}, 461 (1996)];
%
  K.~Funakubo,
  Prog.\ Theor.\ Phys.\  {\bf 96}, 475 (1996);~
%
  A.~Riotto,
  hep-ph/9807454;~
%
 W.~Bernreuther,
 Lect.\ Notes Phys.\  {\bf 591}, 237 (2002);~
%
  J.M.~Cline,
  arXiv:hep-ph/0609145;~
%
  D.E.~Morrissey and M.J.~Ramsey-Musolf,
  New J.\ Phys.\  {\bf 14}, 125003 (2012);~
  T.~Konstandin,
  Phys.\ Usp.\  {\bf 56}, 747 (2013).

\bibitem{D.Delgove}
Talk given by D. Delgove at Double Higgs Production at Colliders Workshop, September 4-9, 2018, Fermilab, USA.        

\bibitem{Fujii:2015jha} 
  K.~Fujii {\it et al.},
  arXiv:1506.05992 [hep-ex].

\bibitem{3h_100TeV}
  D.~Gonçalves, T.~Han, F.~Kling, T.~Plehn and M.~Takeuchi,
  Phys.\ Rev.\ D {\bf 97}, no. 11, 113004 (2018);~
  %
  J.~Chang, K.~Cheung, J.~S.~Lee, C.~T.~Lu and J.~Park,
  arXiv:1804.07130 [hep-ph].


\bibitem{lam3h_2L}
  M.~Brucherseifer, R.~Gavin and M.~Spira,
  Phys.\ Rev.\ D {\bf 90}, no. 11, 117701 (2014);~
%
  M.~Mühlleitner, D.~T.~Nhung and H.~Ziesche,
  JHEP {\bf 1512}, 034 (2015).
  
\bibitem{Coleman:1973jx} 
  S.~R.~Coleman and E.~J.~Weinberg,
  Phys.\ Rev.\ D {\bf 7}, 1888 (1973).

\bibitem{Ford:1992pn} 
  C.~Ford, I.~Jack and D.~R.~T.~Jones,
  Nucl.\ Phys.\ B {\bf 387}, 373 (1992)
  Erratum: [Nucl.\ Phys.\ B {\bf 504}, 551 (1997)].

\bibitem{Hollik:2001px} 
  W.~Hollik and S.~Penaranda,
  Eur.\ Phys.\ J.\ C {\bf 23}, 163 (2002).
  
\bibitem{Tanabashi:2018oca} 
  M.~Tanabashi {\it et al.} [Particle Data Group],
  Phys.\ Rev.\ D {\bf 98}, no. 3, 030001 (2018).
  
\bibitem{Degrassi:2012ry} 
  G.~Degrassi, S.~Di Vita, J.~Elias-Miro, J.~R.~Espinosa, G.~F.~Giudice, G.~Isidori and A.~Strumia,
  JHEP {\bf 1208}, 098 (2012).

\bibitem{ES}
E.~Senaha, in preparation.  
  
\bibitem{Kanemura:2017gbi} 
  S.~Kanemura, M.~Kikuchi, K.~Sakurai and K.~Yagyu,
  Comput.\ Phys.\ Commun.\  {\bf 233}, 134 (2018).

\bibitem{Barbieri:2006dq} 
  R.~Barbieri, L.~J.~Hall and V.~S.~Rychkov,
  Phys.\ Rev.\ D {\bf 74}, 015007 (2006).
%

\bibitem{IDM}
  E.~Lundstrom, M.~Gustafsson and J.~Edsjo,
  Phys.\ Rev.\ D {\bf 79}, 035013 (2009);~
%
  L.~Lopez Honorez and C.~E.~Yaguna,
  JHEP {\bf 1009}, 046 (2010);~
%
  A.~Goudelis, B.~Herrmann and O.~Stål,
  JHEP {\bf 1309}, 106 (2013);~

%
  A.~Arhrib, Y.~L.~S.~Tsai, Q.~Yuan and T.~C.~Yuan,
  JCAP {\bf 1406}, 030 (2014);~
%
  T.~Abe and R.~Sato,
  JHEP {\bf 1503}, 109 (2015);~
 %
  B.~Swiezewska,
  JHEP {\bf 1507}, 118 (2015);~
%
  A.~Arhrib, R.~Benbrik, J.~El Falaki and A.~Jueid,
  JHEP {\bf 1512}, 007 (2015);~
%
  P.~M.~Ferreira and B.~Swiezewska,
  JHEP {\bf 1604}, 099 (2016);~
%
  N.~Blinov, J.~Kozaczuk, D.~E.~Morrissey and A.~de la Puente,
  Phys.\ Rev.\ D {\bf 93}, no. 3, 035020 (2016);~
 %
  P.~Poulose, S.~Sahoo and K.~Sridhar,
  Phys.\ Lett.\ B {\bf 765}, 300 (2017);~
%
  A.~Datta, N.~Ganguly, N.~Khan and S.~Rakshit,
  Phys.\ Rev.\ D {\bf 95}, no. 1, 015017 (2017);~
%
  M.~Hashemi and S.~Najjari,
  Eur.\ Phys.\ J.\ C {\bf 77}, no. 9, 592 (2017).

\bibitem{Kanemura:2016sos} 
  S.~Kanemura, M.~Kikuchi and K.~Sakurai,
  Phys.\ Rev.\ D {\bf 94}, no. 11, 115011 (2016).

\bibitem{Belyaev:2016lok} 
  A.~Belyaev, G.~Cacciapaglia, I.~P.~Ivanov, F.~Rojas-Abatte and M.~Thomas,
  Phys.\ Rev.\ D {\bf 97}, no. 3, 035011 (2018).

\bibitem{IDM_EWPT}
  D.~Borah and J.~M.~Cline,
  Phys.\ Rev.\ D {\bf 86}, 055001 (2012);~
%
  G.~Gil, P.~Chankowski and M.~Krawczyk,
  Phys.\ Lett.\ B {\bf 717}, 396 (2012);~
%
  F.~P.~Huang and J.~H.~Yu,
  arXiv:1704.04201 [hep-ph].
%
\bibitem{Blinov:2015vma} 
  N.~Blinov, S.~Profumo and T.~Stefaniak,
  JCAP {\bf 1507}, no. 07, 028 (2015).
    
\bibitem{Laine:2017hdk} 
  M.~Laine, M.~Meyer and G.~Nardini,
  Nucl.\ Phys.\ B {\bf 920}, 565 (2017).
    
\bibitem{micromegas}
  G.~Belanger, F.~Boudjema, A.~Pukhov and A.~Semenov,
  Comput.\ Phys.\ Commun.\  {\bf 176}, 367 (2007);~
%
  G.~Belanger, F.~Boudjema, A.~Pukhov and A.~Semenov,
  Comput.\ Phys.\ Commun.\  {\bf 185}, 960 (2014).

\bibitem{Aprile:2018dbl} 
  E.~Aprile {\it et al.} [XENON Collaboration],
  Phys.\ Rev.\ Lett.\  {\bf 121}, no. 11, 111302 (2018).

\bibitem{Aaboud:2018xdt} 
  M.~Aaboud {\it et al.} [ATLAS Collaboration],
  Phys.\ Rev.\ D {\bf 98}, 052005 (2018).

\bibitem{Sirunyan:2018ouh} 
  A.~M.~Sirunyan {\it et al.} [CMS Collaboration],
  JHEP {\bf 1811}, 185 (2018).

\bibitem{Branco:2011iw} 
  G.~C.~Branco, P.~M.~Ferreira, L.~Lavoura, M.~N.~Rebelo, M.~Sher and J.~P.~Silva,
  Phys.\ Rept.\  {\bf 516}, 1 (2012).

\bibitem{Dolan:1973qd} 
  L.~Dolan and R.~Jackiw,
  Phys.\ Rev.\ D {\bf 9}, 3320 (1974).

\bibitem{Fuyuto:2014yia} 
  K.~Fuyuto and E.~Senaha,
  Phys.\ Rev.\ D {\bf 90}, no. 1, 015015 (2014).
    
\bibitem{Arnold:1992rz} 
  P.~B.~Arnold and O.~Espinosa,
  Phys.\ Rev.\ D {\bf 47}, 3546 (1993)
  Erratum: [Phys.\ Rev.\ D {\bf 50}, 6662 (1994)].
      
\bibitem{VeffT2L}  
  J.~E.~Bagnasco and M.~Dine,
  Phys.\ Lett.\ B {\bf 303}, 308 (1993);~
%
  J.~R.~Espinosa,
  Nucl.\ Phys.\ B {\bf 475}, 273 (1996).

\bibitem{Tresum}
  R.~R.~Parwani,
  Phys.\ Rev.\ D {\bf 45}, 4695 (1992)
  Erratum: [Phys.\ Rev.\ D {\bf 48}, 5965 (1993)];~
%
  S.~Chiku and T.~Hatsuda,
  Phys.\ Rev.\ D {\bf 58}, 076001 (1998).
 
\bibitem{Funakubo:2012qc} 
  K.~Funakubo and E.~Senaha,
  Phys.\ Rev.\ D {\bf 87}, no. 5, 054003 (2013).

\bibitem{Kainulainen:2019kyp} 
  K.~Kainulainen, V.~Keus, L.~Niemi, K.~Rummukainen, T.~V.~I.~Tenkanen and V.~Vaskonen,
  arXiv:1904.01329 [hep-ph].

\bibitem{Braathen:2019pxr} 
  J.~Braathen and S.~Kanemura,
  arXiv:1903.05417 [hep-ph].

\end{thebibliography}
\end{document}